\documentclass[12pt]{article}
\input epsf.tex

\newcommand{\beq}{\begin{equation}}
\newcommand{\eeq}{\end{equation}}
\newcommand{\beqa}{\begin{eqnarray}}
\newcommand{\eeqa}{\end{eqnarray}}

\newcommand{\tr}{\mathop{\rm Tr}}

\newcommand{\cO}{{\mathcal{O}}}
\newcommand{\cN}{{\mathcal{N}}}

\newcommand{\Gc}{\ensuremath{\Gamma_{\mathrm{cusp}}}} %Cusp anomalous dimension
\newcommand{\bGc}{\ensuremath{\bar{\Gamma}_{\mathrm{cusp}}}} %Cusp anomalous dimension
\newcommand{\gN}{\ensuremath{\sqrt{g_sN}}}            % sqrt of 't Hooft coupling 

\begin{document}

\begin{titlepage}

\begin{flushright}
hep-th/0210115
\end{flushright}

\vspace{1cm}

\begin{center}
{\Large\bf A note on twist two operators in $\cN=4$ SYM and \\
\smallskip
Wilson loops in Minkowski signature
 }

\end{center}
\vspace{3mm}

\begin{center}

Mart\'\i n Kruczenski\footnote[1]{e-mail: \tt martink@physics.utoronto.ca}

\vspace{5mm}

  Department of Physics, University of Toronto\\
       60 St. George St., Toronto, Ontario M5S 1A7, Canada \\

\vspace{3mm}
 
and

\vspace{3mm}
  Perimeter Institute for Theoretical Physics \\
       35 King St. N., Waterloo, Ontario N2J 2W9, Canada \\

\end{center}

\vspace{5mm}

\begin{center}
{\large \bf Abstract}
\end{center}
\noindent

Recently, the anomalous dimension of twist two operators in $\cN=4$ SYM theory
was computed by Gubser, Klebanov and Polyakov in the limit of large
't Hooft coupling using semi-classical rotating
strings in $AdS_5$. Here we reproduce their results for large angular 
momentum by using the cusp
anomaly of Wilson loops in Minkowski signature also computed within the 
AdS/CFT correspondence. In our case the anomalous dimension is related
to an Euclidean world-sheet whose properties are completely determined by the 
symmetries of the problem. This gives support to the proposed identification 
of rotating strings and twist two operators.

\vfill
\begin{flushleft}
October 2002
\end{flushleft}
\end{titlepage}
\newpage

%%%% INTRODUCTION
\section{Introduction}

 The AdS/CFT correspondence \cite{malda,Gubser:1998bc,Witten:1998qj} 
provides a description of the large N limit of
$\cN=4$ SU(N) SYM theory as type IIB string theory on $AdS_5\times S_5$. This
has led to much progress in the understanding of SYM theories in the 't Hooft 
limit \cite{magoo}.
However the precise correspondence between states in $AdS_5\times S_5$ and operators in 
the SYM theory is not known in general. Most 
developments\footnote{See \cite{magoo} for a review and a complete set of references.} are based on the relation between 
chiral primary operators in the SYM theory and supergravity modes in $AdS_5\times S_5$, 
as well as in the relation between Wilson loop operators and
semiclassical world-sheets in $AdS_5\times S_5$.
 In \cite{Berenstein:2002jq}, Berenstein, Maldacena and Nastase extended the 
correspondence to a set of operators in the SYM 
theory corresponding to excited string states in $AdS_5\times S_5$ with
large angular momentum along the $S_5$. In a related development, Gubser, Klebanov and 
Polyakov \cite{Gubser:2002tv}\footnote{See also the recent works \cite{Frolov:2002av,Russo:2002sr}.} 
suggested that states with angular momentum in the $AdS_5$ part should 
correspond to twist two operators in the SYM theory. In free field theory these are
the operators with lowest conformal dimension for a given spin $S$ one example
being:
\beqa
\cO_{\{\mu_1\cdots\mu_S\}} &=& \tr \Phi^I \nabla_{\{\mu_1} \cdots \nabla_{\mu_S\}} \Phi^I, 
%\cO_{(\mu_1\cdots\mu_S)} &=& \tr \Psi \nabla_{(\mu_1} \cdots \nabla_{\mu_S)} \Psi^I \\
%\cO_{(\mu_1\cdots\mu_S)} &=& \tr \Phi_{\alpha\beta} \nabla_{(\mu_1} \cdots \nabla_{\mu_S)} F^{\alpha\beta} \\
\label{eq:t2op}
\eeqa
where $\Phi^I$
%$\Psi$ and $F_{\alpha\beta}$  
are the scalars
% fermions and field strength
of the $\cN=4$ theory, and the indices $\{\mu_1\cdots\mu_S\}$ are symmetrized and traces
removed so that the operator carries spin $S$. The conformal dimension in free field
theory is $\Delta_S=S+2$, and the twist $\Delta_S-S=2$. Other twist two operators can be
constructed using the gauge field or the fermions instead of the scalar fields. The interest
of these operators is that in QCD they give the leading contribution to deep inelastic scattering \cite{DIS}.
The proposal of \cite{Gubser:2002tv} is that such operators can be described, in the 
dual supergravity picture, as macroscopic rotating strings whose energy turns out to be,
 for large $S$, 
\beq
 E \simeq S + \frac{\sqrt{g_sN}}{\pi} \ln S, \ \ \ \ \ (S\rightarrow\infty).
\label{eq:EmS}
\eeq 
In the CFT, $E$ is interpreted as the conformal dimension $\Delta_S$ of the 
corresponding operator, and in fact, in the field 
theory, a similar behavior is obtained at one loop for the operators (\ref{eq:t2op}) 
except that the coefficient in front of $\ln S$ is linear in $g_s N$, where 
$g_s=g_{\mathrm{YM}}^2$ \cite{Gubser:2002tv}. The field theory result is obtained by 
evaluating the expectation value
\beq
\Gamma_{\cO_s}^{[2]} = \langle p | \cO_s | p \rangle = 
C_S (ip_\mu\Delta^\mu)^S \left(\frac{\Lambda}{\mu}\right)^{\gamma_S}, 
\label{eq:GfOs}
\eeq
where $|p\rangle$ is a one particle $\Phi^I$ state with momentum $p$, $\Lambda$ and $\mu$ are
UV and IR cut offs and $\gamma_S$ is the anomalous dimension ($\Delta_S=S+2+\gamma_S$).
 The operator $\cO_s$ is defined as
\beq
\cO_S(\Delta) = \tr \left(\Phi^I \nabla_{\mu_1} \cdots \nabla_{\mu_S} \Phi^I \right) \Delta^{\mu_1}\cdots \Delta^{\mu_S},
\label{eq:OsD}
\eeq
 where $\Delta^\mu$ is an arbitrary vector which is taken to be light-like 
so that only the  traceless part contributes to $\cO_s$. The Green function in eq.(\ref{eq:GfOs})
can be evaluated perturbatively as in \cite{DIS} and the result is that, at one loop, the leading contribution 
(for $S\rightarrow\infty$) to the anomalous dimension comes from the diagram in fig. 1a. which gives the logarithmic dependence 
$\Delta_S = S +2 + \gamma_S$, $\gamma_S\sim \ln S$, $(S\rightarrow\infty)$. This behavior persist at two loops
in supersymmetric and not supersymmetric theories\footnote{See \cite{Axenides:2002zf} for a review of these results and
references to the extensive literature in the subject.}.   

\begin{figure}
\centerline{\epsfxsize=15cm\epsfbox{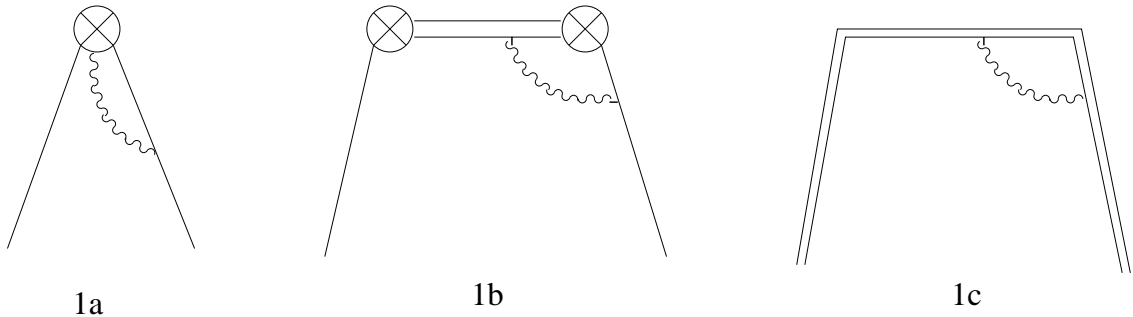}}
\caption{\small Different ways to compute, at 1-loop, the anomalous dimension of the twist two operators in field theory for
large spin. The straight lines indicate scalar propagators, wavy lines denote gluons and double lines 
a Wilson line in the adjoint representation. 1c) is just a Wilson line which can also be computed using the AdS/CFT.}
\end{figure}

In \cite{Korchemsky}, it was observed that the anomalous dimension, for large $S$, 
can also be computed by using Wilson loops.
More precisely, that it is related to the anomalous dimensions of Wilson loops with cusps. This suggests an alternative
method to use the AdS/CFT correspondence to perform the calculation which is to compute the relevant Wilson loop by using
the results of \cite{Maldacena:1998im}. In this paper we pursue this idea an obtain exactly the same result as with 
the rotating string, giving support to the identification between the operators $\cO_S$ and the rotating strings.

In the next section we review the calculation of \cite{Gubser:2002tv} using rotating 
strings an explain how it can be done in an alternative way by using Wilson loops. In 
section 3 we perform the calculation of the Wilson loop by doing an analytic
continuation of previous results in Euclidean AdS space \cite{Drukker:1999zq}. In section 4 we analyze these 
results and give its interpretation in terms
of an Euclidean word-sheet in $AdS_5\times S_5$. In particular we show that the
relevant world-sheet is completely determined by symmetry considerations. Finally, 
we give our conclusions in section 5. 

\section{Rotating strings and Wilson loops}

In this section, after reviewing the calculation of the string rotating in $AdS_5$ \cite{Gubser:2002tv} we 
summarize the argument of Korchemsky and Marchesini \cite{Korchemsky} which suggests 
an alternative computation using Wilson loops.

In~\cite{Gubser:2002tv} it was proposed that twist two operators corresponded to 
(semiclassical) rotating strings in $AdS_5$.
The $AdS_5$ metric in global coordinates (see Appendix) is
\beq
ds^2 = - g_{tt} dt^2 + d\rho^2 + g_{\phi\phi} d\Omega_3^2,
\label{eq:AdSmgc}
\eeq
with $g_{tt}=\cosh^2\rho$ and $g_{\phi\phi} = \sinh^2\rho$. A rotating string is
given by $\rho=\sigma$, $t=\tau$ and $\phi=\omega\tau$ where $\phi$ is an angle
around a maximal circle in the $S^3$. Using the Nambu-Goto action it is easy to compute
the energy and angular momentum as functions of the parameter $\omega$. In the limit 
$S\rightarrow \infty$ we have $\omega\rightarrow 1$ and 
\beqa
P_{+} &=& E+\omega S = 4 \frac{R^2}{2\pi\alpha'}\int_0^{\rho_0} d\rho 
\frac{g_{tt}+\omega^2 g_{\phi\phi}}{\sqrt{g_{tt}-\omega^2 g_{\phi\phi}}} 
\sim \int^{\rho_0} d\!\rho e^{2\rho} \sim e^{2\rho_0}, 
\label{Pp} \\
P_{-} &=& E-\omega S = 4 \frac{R^2}{2\pi\alpha'} 
          \int_0^{\rho_0} d\rho \sqrt{g_{tt}-\omega^2 g_{\phi\phi}}
          \simeq 4\frac{R^2}{2\pi\alpha'}\int_0^{\rho_0} d\rho = 
          \frac{R^2}{\pi\alpha'} 2\rho_0,
\label{Pm}
\eeqa
where $\rho_0$ is such that $\omega = g_{tt}(\rho_0)/g_{\phi\phi}(\rho_0)$ and
consequently $\rho_0\rightarrow \infty$ as $\omega\rightarrow 1$. The approximations
in (\ref{Pp}), (\ref{Pm}) are valid in that limit and, the one in (\ref{Pp}) only up to 
a constant factor. From (\ref{Pp}),(\ref{Pm}), and using that $R^2/\alpha' = \sqrt{g_sN}$, 
we can derive (\ref{eq:EmS}) which, within this approximation can be written as
\beq
P_- \simeq \frac{\gN}{\pi} \ln P_+ .
\label{eq:Pppm}
\eeq
 The approximate results can be thought of as arising from a string in a metric
\beq
ds^2 = e^{2\rho} dx_+ dx_- + dx_-^2 + d\rho^2 + dy_{[2]}^2, \ \ \ \ \ 0<\rho<\rho_0,
\eeq
and extending along $\rho$ and $x_+$. This metric is invariant under translations 
of $x_-$ and under a combined translation in $\rho$ and rescaling in $x_+$. This fact
results in all segments $d\!\rho$ of the string contributing equally to $P_-$ (in (\ref{Pm}))
but weighted by $e^{2\rho}$ to $P_+$ (in (\ref{Pp})), resulting in $P_-\sim \ln P_+$. According to
the rules of the AdS/CFT correspondence, the integrals in $\rho$ should be interpreted in the
field theory as sums of contributions from different scales and it would be interesting to see
if  their behavior can be derived by field theory considerations alone.  

 An alternative approach to computing the anomalous dimension is suggested by the
fact that the operators $\cO_S(\Delta)$ in eq.(\ref{eq:OsD}) arise in a powers series expansion of
\beq
W_{\Delta^\mu} = \tr \left(\Phi^I(\Delta^\mu) e^{\int_0^{\Delta} A_\mu(t\Delta^\mu)\Delta^\mu dt} \Phi^I(0) \right)
 = \sum_{S=0}^{\infty} \frac{1}{S!}\cO_S(\Delta),  
\eeq
 where the Wilson line is in the adjoint representation ($A_\mu=A^a_\mu C_{abc}$). 
Again we can compute
\beq
\langle p | W({\Delta^\mu}) | p \rangle = \sum \frac{1}{S!} 
C_S (ip_\mu\Delta^\mu)^S \left(\frac{\Lambda}{\mu}\right)^{\gamma_S}.
%-i e^{ip_\mu\Delta^{\mu}} 
%\int_0^1 \frac{dt}{t} \left(e^{-it(p\Delta)}-1\right) (p^2)^{2-d/2} \Gamma(d/2-2).
\label{eq:Wdr}
\eeq 
Thus, the coefficient of $(p\Delta)^S$ in a power series expansion of 
$\langle p | W({\Delta^\mu}) | p \rangle$ in $(p\Delta)$ 
determines the anomalous dimension of the operator $\cO_S$. For example, a 1-loop
calculation as in fig.1b gives the same result for $\gamma_S$ as the one obtained
from fig.1a.  In particular, the large $S$ behavior of $\gamma_S$ is related to
the large $p\Delta$ behavior of $\langle p | W({\Delta^\mu}) | p \rangle$. 

 It was argued in \cite{Korchemsky} that, in this limit, the relevant contribution to the anomalous
dimension comes from soft gluons\footnote{See \cite{Collins} 
for a detailed description of these ideas. We thank G. Korchemsky for clarifying this point to us.}. 
This fact allows the use of the eikonal approximation for the propagator of the external particles, in 
this case $\Phi$. In Feynman diagrams, this amounts to replace the propagator of $\Phi$ with momentum $p-k$ by:
\beq
\frac{1}{(p-k)^2+i\epsilon} \rightarrow \frac{1}{-2p . k+i\epsilon},
\eeq
where $p$ is the momentum of the external $\Phi$ line ($p^2=0$) and $k$ the momentum 
of an attached gluon line. 

The calculation is then reduced to the computation of a Wilson line ($W_M$) as the one in fig.1c,
where the $\Phi$ propagators are replaced by Wilson lines (in the adjoint 
representation) going from infinity to 0 and from $x^\mu=\Delta^\mu$ to infinity
with momentum $p$ and $-p$, respectively. These are joined by a Wilson line
going from $0$ to $x^\mu=\Delta^\mu$ which is the same as the one appearing in
fig. 1b. The precise relation is that
\beq
 \langle p | W({\Delta^\mu}) | p \rangle \simeq e^{ip.\Delta} W_M,
\eeq
where $W_M$ is the Wilson line previously discussed. 

 The Wilson line has an anomalous dimension due to the fact that it possesses cusps \cite{Polyakov}.
As explained in \cite{Korchemsky} there is a subtlety because the expectation value of 
the Wilson loop diverges when part of it is light-like. If we regulate the Wilson loop by taking
$\Delta^2$ and $p^2$ non-vanishing, we expect, from \cite{Polyakov}, each cusp to contribute with
\beq
 W_{\mathrm{cusp}} \sim \left(\frac{L}{\epsilon}\right)^{-\Gc(\gamma)},
\eeq
where $L$ and $\epsilon$ are IR and UV cut offs, and the anomalous dimension $\Gc(\gamma)$ is
only a function of the angle between $p$ and $\Delta$ defined as 
\beq
\cosh \gamma = \frac{p.\Delta}{\sqrt{p^2\Delta^2}}.
\eeq
 Since we are interested in the region where $p.\Delta$ is very large we should compute $\Gc(\gamma)$ for
large values of $\gamma$. It was shown in \cite{gcusp} that, in that limit, the anomalous dimension behaves as
\beq
\Gc(\gamma) \simeq \bGc |\gamma| \simeq \bGc \ln(p.\Delta), \ \ \ \ (\gamma\rightarrow\infty),
\label{eq:Gcbehaviour}
\eeq
where $\bGc$ is a constant that depends only on $g_{\mathrm{YM}}$ and $N$. In the last approximation we kept only
the leading contribution for large $(p.\Delta)$ which is independent of $\Delta^2$ and $p^2$.   
 Considering the contribution of both cusps we obtain that   
\beq
\langle p | W({\Delta^\mu}) | p \rangle \sim e^{ip.\Delta} \left(\frac{L}{\epsilon}\right)^{-2\bGc \ln (p. \Delta)}
=  e^{ip.\Delta} (p. \Delta)^{-2\bGc \ln \left(\frac{L}{\epsilon}\right)},
\eeq
where the approximation is valid for large $(p.\Delta)$. The fact that 
\beq
\sum_{n=0}^{\infty} a_n \frac{x^n}{n!} \simeq x^\alpha e^x,\ \ \ \ (x\rightarrow\infty)\ \ \ \ \mathrm{if}\ \ \ 
a_n\sim n^{\alpha},\ \ \ (n\rightarrow\infty),
\label{eq:identity}
\eeq
allows us to conclude that 
\beq
\langle p | W({\Delta^\mu}) | p \rangle \sim
 \sum_S \frac{1}{S!} 
S^{-2\bGc \ln \left(\frac{L}{\epsilon}\right)}(ip\Delta)^S = 
 \sum_S \frac{1}{S!} 
\left(\frac{L}{\epsilon}\right)^{-2\bGc \ln S}(ip\Delta)^S .
\label{eq:andim}
\eeq
The result (\ref{eq:identity}) can be derived by showing that, if we define a function $f(x)$ through
\beq
a_n = \int_0^1 s^n f(s) ds ,
\eeq
then both, the behavior of $a_n$ for $n\rightarrow\infty$ and the behavior of 
\beq
e^{-x} \sum a_n \frac{x^n}{n!} = \int_0^1 e^{(s-1)x} f(s) ds ,
\eeq
for $x\rightarrow\infty$, are determined by the properties of the function $f(s)$ near $s=1$. 

 Finally, comparing eq.(\ref{eq:andim}) with eq.(\ref{eq:Wdr}) we see that, after identifying $\Lambda/\mu\sim L/\epsilon$,  
the anomalous dimension of the operator $\cO_S$ is equal to
\beq
\Delta_S - S -2 = \gamma_S \simeq -2\bGc \ln S,\ \ \ \ (S\rightarrow\infty).
\label{eq:adfWl}
\eeq
 The above argument ignores the problem of setting $\Delta^2=p^2=0$ since in that case the terms discarded
in eq. (\ref{eq:Gcbehaviour}) are singular. We refer the reader to the original reference \cite{Korchemsky} for a more
rigorous treatment which gives the same result (\ref{eq:adfWl})\footnote{In \cite{Korchemsky} the result was
equivalently expressed in terms of the splitting function $P(z)$.}.

In the next section we obtain $\bGc$ using the AdS/CFT correspondence to compute a Wilson loop with a cusp
in Minkowski signature. This is simply an analytic continuation of the results of \cite{Drukker:1999zq} where
the calculation was done in Euclidean space. 
 However, before that, we conclude this section by reviewing some properties of
the Wilson loop cusp anomaly at one loop which will be useful to compare with the AdS/CFT result.

 For the $\cN=4$ case the one loop anomaly can be taken from \cite{Drukker:1999zq}, but its behavior
with the angle is similar to the one in QCD \cite{Polyakov}. If the cusp has an 
 angle $\Theta$ (where $\Theta=\pi$ defines a straight line) the Wilson loop behaves as
\beqa
W_E &=& \left({L\over\epsilon} \right)^{-\Gc(\Theta)} , \\
\Gc(\Theta) &=& \frac{g_sN}{4\pi^2}\left((\pi-\Theta)\cot\Theta+1\right) .
\label{eq:1loop}
\eeqa
 The result (\ref{eq:1loop}) can be continued to Minkowski space by replacing 
$\Theta\rightarrow\pi+i\gamma$. The resulting $\Gc^{(M)}(\gamma)=\Gc(\pi+i\gamma)$ satisfies:
\beqa
\Gc^{(M)}(\gamma) &\simeq& -\frac{g_sN}{6\pi^2} \gamma^2 , \ \ \ (\gamma \rightarrow 0), \\ 
\Gc^{(M)}(\gamma) &\simeq& -\frac{g_sN}{4\pi^2} \gamma   , \ \ \  (\gamma \rightarrow \infty), \\ 
\eeqa
which then gives at 1-loop: $\bGc=-(g_sN)/4\pi^2$.

As an aside note that the Euclidean result has the property 
\beq
\Gc(\Theta) \simeq \frac{g_sN}{4\pi^2} \frac{1}{\Theta}, \ \ \ \   (\Theta\rightarrow 0), 
\label{eq:1oT1loop}
\eeq
which can be understood from conformal invariance. In fact
the flat metric
\beq
ds^2 = dr^2 + r^2 d\theta^2,
\eeq
is invariant (up to a conformal factor) under the transformation 
$r\rightarrow r^\lambda $, $\theta\rightarrow \lambda\theta$. This alters 
the periodicity of $\theta$ and so, is not actually a conformal transformation. However, in
the limit $\Theta\rightarrow 0$ we need to consider only small values of $\theta$ and
the periodicity can be ignored implying that 
\beq
W\sim \left({L\over \epsilon}  \right)^{-\Gc(\Theta)}  ,
\label{eq:1oT}
\eeq
is invariant only if $\Gc(\Theta)\sim 1/\Theta$, as obtained in (\ref{eq:1oT1loop}).

%%%%%%SECTION 3
\section{Cusp anomaly from supergravity.}
\label{sec3}

The cusp anomaly can be computed by means of the AdS/CFT correspondence
by computing the expectation value of a Wilson loop with a cusp following
\cite{Maldacena:1998im}. This computation was done in \cite{Drukker:1999zq} in the Euclidean case. 

In this section we summarize their calculation and perform an analytic 
continuation to Minkowski signature which allows us to compute $\bGc$ and, by
plugging the result into (\ref{eq:adfWl}), to reproduce (\ref{eq:EmS}).

In the next section we reobtain these
results by a direct calculation and obtain the shape of the world-sheet
ending in the boundary of AdS. One
extra consideration is that here we need a Wilson loop in the adjoint
representation. This corresponds to a quark and anti-quark propagating along 
the loop. We will consider simply a double cover of the surface,
i.e. we include an extra factor of 2 at the end. Another point is that the Maldacena
correspondence allows us only to compute Wilson lines of the type
\beq
W = \exp\left(i\int A_{\mu} \dot{x}^{\mu} + \Phi^I \theta_I |\dot{x}|\right).
\eeq
For the section parallel to $\Delta^{\mu}$ there is no problem since $|\Delta|=0$. 
For the lines representing the incoming particles, instead of the operator (\ref{eq:t2op}) 
we will be computing the anomalous dimension of an operator which is a linear combination 
of that one and the one obtained by replacing $\Phi$ by the gauge field. All previous considerations
still apply since (\ref{eq:t2op}) was taken only as an example. It will be interesting to further analyze
this issue and see if supergravity can distinguish between the different twist two operators, or
if it predicts that all of them have the same anomalous dimension for large $S$ and large $g_s N$.

\begin{figure}
\centerline{\epsfxsize=7cm\epsfbox{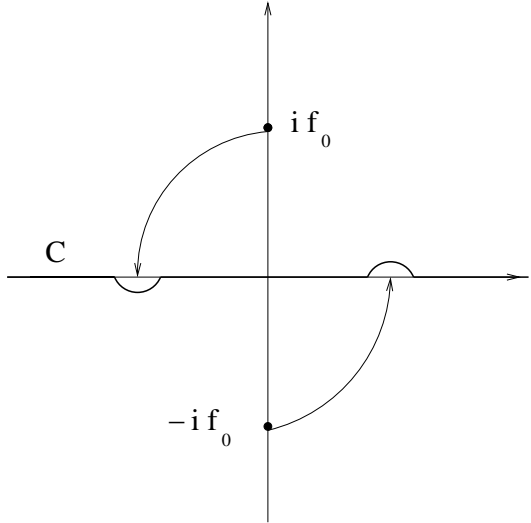}}
\caption{\small Rotation of $f_0$ in the complex plane necessary to get an imaginary angle $\Theta$.}
\label{fig:rot} 
\end{figure}

After these preliminary considerations, lets go back to the computation of the Wilson loop, 
starting by the Euclidean case of \cite{Drukker:1999zq}. Consider Euclidean AdS space in 
Poincare coordinates where the metric is
\beq
ds^2=\frac{1}{z^2}\left(dz^2+dx_1^2+dx_2^2+dx_3^2+dx_4^2\right).
\eeq
Taking polar coordinates ($\rho$,$\theta$) in the plane ($x_1$,$x_2$)
we will compute the area of a world-sheet which ends in the boundary $z=0$ on two half-lines
given by $\theta=\pm \Theta/2$ and $x_1>0$. By scale invariance the equation of the surface
can be written as 
\beq
z = \frac{\rho}{f(\theta)},
\eeq
with $f(\pm \Theta/2) = \infty$. Also, by symmetry, we should have that
$\frac{df}{d\theta}(\theta=0)=0$. Extremizing the Nambu-Goto action for the
string with respect to the function $f(\theta)$ we obtain the desired result
(more details can be found in \cite{Drukker:1999zq} or in the next section):
\beqa
\Theta &=& \int_{-\infty}^{+\infty} 
 \frac{f_0\sqrt{1+f_0^2}}{(u^2+f_0^2)\sqrt{(1+u^2+f_0^2)(1+z^2+2f_0^2)}} du,
\label{eq:Tf0E}\\
A &=& \frac{R^2}{2\pi\alpha'}\ln \frac{L}{\epsilon} \int_{-\infty}^{+\infty} 
      \left\{\sqrt{\frac{1+u^2+f_0^2}{1+u^2+2f_0^2}}-1\right\}.  
\eeqa
Here $A$, is the desired area of the world sheet, given as a function of the angle
$\Theta$ implicitly through the parameter $f_0$. The term $-1$ inside the 
integral is an infinite subtraction constant which makes the area 
finite. Also, $L$ and $\epsilon$ are an infrared and ultraviolet cut-offs 
such that $\epsilon<\rho<L$. According to \cite{Maldacena:1998im}, $\ln W_M$ is equal 
to the area $A$ and so, we identify the cusp anomaly as
\beq
\Gamma_{\mathrm{cusp}}(\Theta) = -\frac{R^2}{2\pi\alpha'}\int_{-\infty}^{+\infty} 
      \left\{\sqrt{\frac{1+u^2+f_0^2}{1+u^2+2f_0^2}}-1\right\}.
\label{eq:Ecusp}
\eeq
For example, as a check, we can see that the limit $f_0\rightarrow\infty$
corresponds to $\Theta\rightarrow 0$ and the cusp anomaly behaves as
\beq
\Gamma_{\mathrm{cusp}}(\Theta) \simeq  -\frac{c^2}{\Theta}, \ \ \ \ (\Theta\rightarrow 0).
\eeq
We see the expected behavior $1/\Theta$ of eqns.(\ref{eq:1oT1loop}),(\ref{eq:1oT}).
The constant $c$ is given by
\beq
c = \int_{-\infty}^{\infty} \frac{du}{(1+u^2)^{3/2}(2+u^2)^{1/2}}.
\eeq
Now we want to do an analytic continuation in $\Gamma_{\mathrm{cusp}}(\Theta)$ 
to $\Theta = \pi + i\gamma$ with $\gamma$ real. It appears, from 
eq.(\ref{eq:Tf0E}), that if we analytically continue $f_0$ to imaginary values,
$\Theta$ becomes purely imaginary. However, a singularity develops
in the integral at $u=f_0$ which gives the extra $\pi$. Indeed, to 
handle the singularities we can perform the analytic 
continuation more carefully as in fig.\ref{fig:rot}. Thus, we obtain a recipe to 
avoid the poles at $u=\pm f_0$. The result is:
\beq
\Theta = i \int_{-\infty}^{+\infty}
 \frac{f_0\sqrt{1-f_0^2}}{(u^2-f_0^2+i\epsilon)\sqrt{(1+u^2-f_0^2)(1+u^2-2f_0^2)}} du.
\eeq
 We can separate the real and imaginary part by doing explicitly the integrals in the small
half circles around the poles obtaining
\beqa
\Theta &=& \pi + i \gamma, \\
\gamma &=& \mathrm{P.P.} \int_{-\infty}^{+\infty}  
\frac{f_0\sqrt{1-f_0^2}}{(u^2-f_0^2)\sqrt{(1+u^2-f_0^2)(1+u^2-2f_0^2)}} du,    
\label{eq:gf0}
\eeqa
where $\gamma$ is real (as long as $|f_0|<1/\sqrt{2}$) and given by the 
principal part of the integral. Finally, from (\ref{eq:Ecusp}),  the cusp anomaly is given by
\beq
\Gamma_{\mathrm{cusp}}(\gamma) = -\sqrt{g_sN} \frac{1}{2\pi}\int_{-\infty}^{+\infty} 
      \left\{\sqrt{\frac{1+u^2-f_0^2}{1+u^2-2f_0^2}}-1\right\}.
\label{eq:Gf0}
\eeq
We can consider first the limit $\gamma\rightarrow 0$ which corresponds to 
$f_0\rightarrow 0$ (note that now $\Theta\rightarrow\pi$ so this limit is 
different from the limit $\Theta\rightarrow 0 $ considered before).
 The result is that
\beq
\begin{array}{lll}
            &\gamma   \simeq  -\pi f_0, \ \ \ \    
\Gc(\gamma) \simeq \gN \frac{1}{4} f_0^2, &\ \ \ (f_0\rightarrow 0)  \\
\Rightarrow &\Gc(\gamma) \simeq \gN\frac{\gamma^2}{4\pi^2}, &\ \ \ (\gamma\rightarrow\infty),
\end{array}
\eeq
which has a similar behavior as the one loop result albeit with a different
coefficient. 

For us it is more interesting the behavior of $\Gamma_{\mathrm{cusp}}(\gamma)$ when
$\gamma\rightarrow\infty$. As we noticed before we need $f_0 <1/\sqrt{2}$. 
We can see easily from the integral in eq.(\ref{eq:gf0}) that the limit 
$f_0\rightarrow 1/\sqrt{2}$ corresponds precisely to 
$|\gamma|\rightarrow\infty$. In this limit we have
\beq
\begin{array}{clclclclcl}
&\gamma &\simeq &  \sqrt{2} \ln \delta, &\ \ \ &   
\Gamma_{\mathrm{cusp}}(\gamma) &\simeq &   \frac{1}{2\pi}\frac{1}{\sqrt{2}}\ln \delta, &  
       & (f_0\rightarrow 1/\sqrt{2}),  \\
\Rightarrow & \Gamma_{\mathrm{cusp}}(\gamma) 
  &\simeq & -\gN\frac{1}{2\pi}\frac{1}{2}|\gamma| & \Rightarrow &  
 \bGc &=& -\gN\frac{1}{4\pi} && (\gamma\rightarrow -\infty)\label{eq:Gcf},
\end{array}
\label{eq:Gcsugra}
\eeq
where $1-2f_0^2=2\delta$ and
the limiting behaviors are obtained by considering the contribution to the 
integrals from the region $u\rightarrow 0$. For example for $\gamma$ we have
\beq
\gamma \simeq 
  \int \frac{du}{(2u^2-1)\sqrt{\frac{1}{2}+u^2}\sqrt{u^2+2\delta}} 
\simeq
  -\sqrt{2}\int_{-\epsilon}^{\epsilon} \frac{du}{\sqrt{u^2+2\delta}} 
= -
  2 \sqrt{2}\mathrm{arcsinh}(\frac{\epsilon}{\sqrt{2\delta}}) 
\simeq 
  \sqrt{2}\ln\delta,
\eeq
Where $\delta\ll\epsilon\ll1$ and we kept only the leading dependence in $\delta$ (for $\delta\rightarrow 0$). 
The calculation for $\Gamma_{\mathrm{cusp}}(\gamma)$ is similar.

Again, the behavior is similar to that of the one loop result but with a 
different coefficient. We can plug the result (\ref{eq:Gcsugra}) into eq.(\ref{eq:adfWl}) obtaining
\beq
\gamma_S = -2\bGc^{\mathrm{(adj.)}} \ln S = \frac{\sqrt{g_sN}}{\pi} \ln S,
\eeq
where, as explained at the beginning of this section, an extra factor of 2 is included, i.e. 
we took $\bGc^{\mathrm{(adj.)}}=2\bGc=-\gN/2\pi$, to account for the fact that the Wilson
loop should be in the adjoint representation. We see that the result agrees with 
that obtained from the rotating string calculation (\ref{eq:EmS}). First,
the linear dependence with $\ln S$ agrees. This follows from the non-trivial fact 
that the Wilson loop in supergravity diverges linearly in $\gamma$ for large $\gamma$.  
 The dependence on the coupling constant also agrees but that was expected since the area
of a Wilson loop scales as $R^2$. On the other hand also the numerical coefficient is the 
same giving support to the identification between twist two operators and rotating strings
in $AdS$ space. In the next section we will see, that in our case, the coefficient follows
from the area of the limiting surface (at $f_0=1/\sqrt{2}$) which can be determined
by symmetry considerations alone. This gives hope that the result can be understood 
directly in the field theory using symmetry arguments.

\section{Wilson loop in Minkowski signature}

In this section we use the AdS/CFT correspondence to compute the cusp 
anomaly directly in Minkowski signature\footnote{See \cite{Tseytlin:2002tr} for another example of 
a calculation of Wilson loops in Minkowski space.}. This will give an interpretation
to eqns. (\ref{eq:gf0}),(\ref{eq:Gf0}) in terms of an Euclidean world-sheet in $AdS_5$. We also
analyze the surface corresponding to the limiting case $f_0=1/\sqrt{2}$ and
show that it can be found using only symmetry arguments. 

\begin{figure}
\centerline{\epsfxsize=5cm\epsfbox{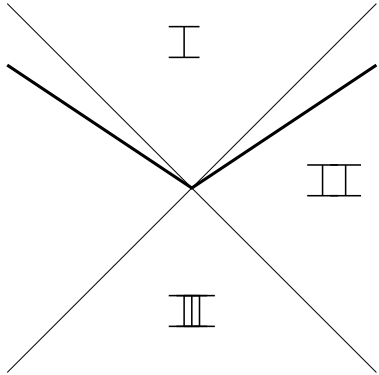}}
\caption{\small Wilson loop in the form of a wedge (thick lines). The
thin lines show the light cone.}
\label{fig:WlM}
\end{figure}

\begin{figure}
\centerline{\epsfxsize=5cm\epsfbox{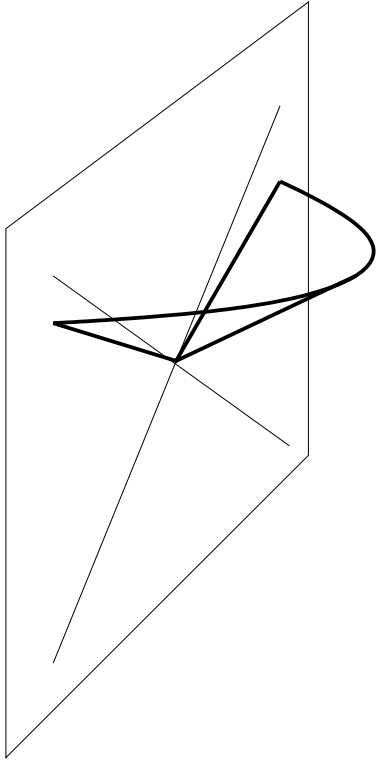}}
\caption{\small Sketch of the surface (in $AdS_5$) used to compute the Wilson loop of figure \ref{fig:WlM}.}
\label{fig:WlMAdS5}
\end{figure}

 Consider a Wilson loop with a cusp as the one depicted in fig. \ref{fig:WlM}. 
In region II, we can use coordinates $(\rho,\xi)$\footnote{These are coordinates in the boundary
of AdS space. $\rho$ should not be confused with the radial direction in AdS space as used before.} defined by
\beqa
x &=& \rho \cosh \xi, \\
t &=& \rho \sinh \xi, 
\eeqa
which simplify the equations of motion.
These coordinates can be extended to region I if we consider $\rho$ to be
purely imaginary and $\xi=i\frac{\pi}{2} + \tilde{\xi}$ with $\tilde{\xi}$ real. 
As before we can use an ansatz
\beq
z = \frac{\rho}{f(\xi)}.
\eeq
Given the above properties of the coordinates ($\rho$,$\xi$) and the fact that
$z$ is real, we need $f(\xi)$ to be purely imaginary when $\xi$ is in region 
I and also to vanish on the light cone ($\xi\rightarrow\infty$), since 
there $\rho$ vanishes. These are complications introduced by the choice of coordinates, 
perhaps at this point it is useful to point out that the resulting surface has a simple 
shape as shown in fig. (\ref{fig:WlMAdS5}). One may think that these complications could have
been avoided by considering a Wilson line contained within region I. However in that case and 
for large angles, there is no surface in AdS space ending on such a Wilson line. 

Since the world-sheet ends perpendicularly to the boundary, it has to be Euclidean. 

 The action for such a world-sheet is given (in region II) by
\beqa
S^{II}_E &=& 
%\int \frac{1}{z^2} 
%\sqrt{\left(\frac{\partial z}{\partial t}\right)^2
%    - \left(\frac{\partial z}{\partial x}\right)^2 -1 } 
%\int \rho d\rho d\xi \sqrt{
% -\frac{1}{\rho^2}\left(\frac{\partial z}{\partial \xi}\right)^2
%+\left(\frac{\partial z}{\partial \rho}\right)^2 -1}   \\
  \int \frac{d\rho}{\rho} d\xi \sqrt{f'^2-f^2-f^4} \\
  &=& \ln \frac{L}{\epsilon} \int d\xi \sqrt{f'^2-f^2-f^4}. 
\eeqa
The fact that the integrand does not depend explicitly on $\xi$ implies
a conservation law:
\beq
\frac{f'^2}{\sqrt{f'^2-f^2-f^4}} - \sqrt{f'^2-f^2-f^4} = C,
\eeq
with $C$ a constant. The solution is given then by
\beq
\xi = \frac{1}{2} \gamma +
  f_0\sqrt{1-f_0^2} \int_{f}^{\infty} 
   \frac{d\!f}{f\sqrt{(1+f^2)(f^2+f_0^2)(1+f^2-f_0^2)}},  
\eeq
where for later convenience we introduce $f_0$ through the relation
\beq
C^2 = f_0^2 - f_0^4 ,\ \ \ f_0^2<1.
\eeq
Furthermore, for $f\rightarrow\infty$, namely $z\rightarrow 0$, we have $\xi\rightarrow\gamma/2$,
i.e. the world-sheet ends on the line $(x,t)= \lambda(\cosh \gamma/2,\sinh \gamma/2)$. Normally, we would
continue the surface until $f'=0$ where we would match with the other half ending on 
$(x,t)= \lambda(-\cosh \gamma/2,\sinh \gamma/2)$. Here, $f'$ can be $0$ only in region I, so we need to
continue the boundary coordinates into that region. First, we observe that,  
at $f=0$, the integral diverges implying that $x/t = \tanh\xi\rightarrow 1$, 
namely, in the boundary,  we approach the light-cone. On the other hand $\rho=\sqrt{x^2-t^2}$ also
vanishes. We can compute the value of $z$ by taking the limit $(\xi\rightarrow\infty)$ with $x$ fixed:
\beq
z = \lim_{\xi\rightarrow \infty} \frac{\rho}{f(\xi)} = 
x \lim_{\xi\rightarrow\infty}\frac{\sqrt{1-\tanh^2\xi}}{2e^Be^{-\xi}}=x
\,e^{-B} = t\,e^{-B},
\eeq
where $B$ is a constant given by
\beq
B = \frac{\gamma}{2} + 
\int_0^\infty \frac{d\!f}{f\sqrt{1+f^2}}\left\{\frac{f_0\sqrt{1-f_0^2}}{\sqrt{(f^2+f_0^2)(1+f^2-f_0^2)}}-1\right\}.
\eeq
To continue beyond the light cone (in the boundary) we have to take $f$ imaginary, then, in region I, 
 $f$ should extend up to $i f_0$ where $f'=0$ to match the other half of the surface. Although 
this may seem strange we want to emphasize again that the surface is perfectly smooth, the analytic
continuation is necessary only because of the coordinates we are using. 
 This is most easily done by defining a variable $u^2 = f^2+f_0^2$ which
extends from $0$ to $\infty$, meaning that $f$ extends from $if_0$ to $0$ along the imaginary axis and from 
$0$ to $\infty$ along the real axis. In this way, and considering also the corresponding formula for the area,
we reproduce the formulas (\ref{eq:gf0}) and (\ref{eq:Gf0}).

It is interesting to study the limiting surface obtained when 
$f_0\rightarrow 1/\sqrt{2}$. In fact in that case $f'=0$ and the surface is simply
\beq
z = \sqrt{2} \rho = \sqrt{2} \sqrt{t^2-x^2}.
\eeq
This surface ends on the light cone $t=\pm x$, $t>0$ and extends only for $t>x$. 
However since $z>\rho$  it is always outside the AdS light cone $z=\rho$. In fact
one can see that a surface $z=\alpha\rho$, $\alpha>1$ has zero area for $\alpha=1$ 
and $\alpha=\infty$ which corresponds to a light cone and two light-like planes
respectively. That implies that the area will have an extreme for an intermediate 
value which is precisely $\alpha=\sqrt{2}$. Note also that the surface cannot be continued
to Euclidean space. That will give $z^2 = -2 (t^2+x^2)$ which has no solutions\footnote{However, one can
also continue $z\rightarrow i z$ and obtain a surface in de Sitter space.}.

It is interesting to study this surface
in global coordinates. If we take coordinates $U,V,W,X,Y,Z$ such that
\beq
U^2+V^2-X^2-Y^2-Z^2-W^2 = R^2,
\eeq
and use the relation between these and Poincare coordinates that we give in the Appendix, 
we can see that the surface is given by
\beq
V^2-X^2 = \frac{R^2}{2},\ \ \ U^2-W^2 = \frac{R^2}{2}, \ \ \ Y=Z=0.
\eeq
This surface is in fact completely determined by the symmetries if one assumes that
it is unique. Indeed, the light cone where the surface ends is invariant under scale 
transformations and boosts in $(x,t)$. This corresponds in global coordinates
to boosts in $(U,W)$ and $(V,X)$, respectively (see Appendix) and implies that the equation
determining the surface should be of the form
\beq
f(U^2-W^2, V^2-X^2) = 0 ,
\label{eq:surfaceeqn}
\eeq
since reflection symmetry in $Y$ and $Z$ implies $Y=0$, $Z=0$. On the other hand, 
$U^2-W^2 + V^2-X^2 = R^2$ and so (\ref{eq:surfaceeqn}) can only be of the form
\beqa
V^2-X^2 &=& a \frac{R^2}{2}, \\
U^2-W^2 &=& (1-a) \frac{R^2}{2}.
\eeqa
Since there is an extra symmetry that interchanges $(V,X)$ with $(U,W)$, unless
$a=1/2$ there would be two different surfaces. Assuming that the surface is unique  
determines $a=1/2$.

Returning to Poincare coordinates, we can see that the induced metric on the surface 
is
\beq
ds^2 = \frac{R^2}{2}\left(\frac{d\rho^2}{\rho^2} + d\xi^2\right),
\eeq
and the area is given by
\beq
A = \frac{R^2}{2\pi\alpha'}\frac{1}{2} \int \frac{d\rho}{\rho} \int_{-\infty}^{+\infty} d\xi,
\eeq
which diverges. However, performing the $\rho$ integral between $\epsilon$ and $L$ as
before and the $\xi$ integral between $-\gamma/2$ and $\gamma/2$ we get
\beq
A = \frac{R^2}{2\pi\alpha'}\frac{1}{2} \gamma \ln \frac{L}{\epsilon},
\eeq
which implies that $\bGc=-\gN /4\pi$ as in (\ref{eq:Gcf}). We see that the value $a=1/2$ is the one that
ultimately determines the anomalous dimension and, as we discussed, is fixed by the 
requirement that the solution be unique.  

%%%%%SECTION 5
\section{Conclusions}
\label{5sec}

 We have computed the anomalous dimension of twist two operators, in $\cN=4$ SYM,
in the limit of large angular momentum and large 't Hooft coupling by using Wilson loops
in Minkowski space. The results are in agreement with the ones of \cite{Gubser:2002tv} 
which where obtained by computing the energy-angular momentum relation of a
semiclassical string rotating in AdS space. The agreement is not surprising
since both calculations are done using the AdS/CFT correspondence but tests
the identification between those operators and the rotating strings. 

 Furthermore the Euclidean world-sheet that determines the value of the Wilson
loop and the anomalous dimension is uniquely determined by the symmetries of the
problem. Those are isometries of AdS space and so, correspond to conformal 
transformations in the SYM theory. It would be interesting if this can be
done directly in the field theory, particularly in view of the fact that certain 
Wilson loops can be computed in $\cN=4$ SYM and agree with the AdS/CFT result \cite{Erickson:2000af}. 
In our case a short calculation suggests that, near a cusp,  summing ladder diagrams as 
in \cite{Erickson:2000af} is not enough to obtain the anomalous dimension.   

\section{Acknowledgments}

 I am very grateful to J. Maldacena, R. Myers, A. Peet, A. Tseytlin, and K. Zarembo for
various comments and suggestions. We also thank G. Korchemsky for correcting a statement
made in the first version of this paper. This research was supported in part by    
NSERC of Canada and Fonds FCAR du Qu\'ebec.

\appendix

\section*{Appendix A}

%The field theory. Scaling. Axial gauge.

In the text various standard coordinates were used to parameterize $AdS_5$. For
completeness we give here the corresponding definitions. Introducing Cartesian
coordinates $(X,Y,Z,W,U,V)$ in $R^{4,2}$, $AdS_5$ is the manifold defined by
the constraint:
\beq
-X^2-Y^2-Z^2-W^2+U^2+V^2 = R^2,
\eeq
for some arbitrary constant $R$. The metric is the one induced by the one in $R^{4,2}$:
\beq
ds^2 = dX^2 + dY^2 + dZ^2 + dW^2 -dU^2 - dV^2.
\eeq
Global coordinates $(\rho,t,\theta_{1,2,3})$ as used in eq.(\ref{eq:AdSmgc}) are defined by  
\beq
\begin{array}{lll}
X=R\sinh\rho\cos\theta_1, & Y = R\sinh\rho\sin\theta_1\cos\theta_2, & Z = R\sinh\rho\sin\theta_1\sin\theta_2\cos\theta_3, \\
W=R\sinh\rho\sin\theta_1\sin\theta_2\sin\theta_3, & U = R\cosh\rho\cos t, & V = R\cosh\rho\sin t. 
\end{array}
\eeq
The metric is given by 
\beq
ds^2 = R^2(-\cosh^2\rho dt^2 + d\rho^2 + \sinh^2\rho d\Omega_{[3]}^2),
\eeq
where $\Omega_{[3]}$ is a 3-sphere parameterized by $\theta_{1,2,3}$. The radial coordinate $\rho$ 
extends from $0$ to $\infty$. It is also useful to introduce another radial coordinate $0<\xi<\pi/2$
through the relation $\cosh\rho=1/\cos\xi$. The metric changes to
\beq
ds^2 = \frac{R^2}{\cos^2\xi} \left(-dt^2 + d\xi^2 + \sin^2\xi d\Omega_{[3]}^2\right).
\eeq
 In these coordinates the surface $V^2-X^2=R^2/2$ discussed in the text is described by the equation
\beq
\sin^2 t = \frac{1}{2} \cos^2\xi + \sin^2\xi \cos^2 \theta_1 .
\eeq
Poincare coordinates $(t,x_{1,2,3},z)$ cover only half of $AdS_5$ and are defined by
\beq
\begin{array}{lll}
X = R\, {x_1/ z}, &  Y = R\, {x_2/ z}, &  Z = R\, {x_3/ z}, \\
W = -{1\over 2z}(-R^2+z^2+x_i^2-t^2), & U = {1\over 2z}(R^2+z^2+x_i^2-t^2), & V=R\,{t/ z},
\end{array}
\eeq  
with a metric
\beq
ds^2 = \frac{R^2}{z^2}\left(dz^2+dx_i^2-dt^2\right) .
\eeq
 In the text we used the fact that a boost in direction $x_1$ corresponds to a boost in directions
$X,V$ and a scale transformation to a boost in directions $W,U$. The first property is obvious
since $U$ and $W$ are invariant under a boost along $x_1$ and the only change is in $X,V$, being precisely the
same boost as in $x_1,t$. To see the second property we have simply to notice that 
\beq
U_+=U+W = R^2/z,\ \ \ U_-=U-W = \frac{1}{z}(z^2+x_i^2-t^2) ,
\eeq
transform as $U_+\rightarrow \lambda U_+$, $U_-\rightarrow U_-/\lambda$ , i.e. a boost, if we 
rescale all coordinates by a factor $\lambda$. Finally the equation of the surface discussed in the text:
\beq
z^2=2(x^2-t^2) ,
\eeq
is equivalent to 
\beq
\frac{R^4}{U_+^2} = 2\left(\frac{V^2R^2}{U_+^2}-\frac{X^2R^2}{U_+^2}\right)\ \ \ \Rightarrow \ \ \ V^2-X^2=\frac{R^2}{2} ,
\eeq
as used in section 4.

\end{document}